\makeatletter
\@ifundefined{@parse@version@dash}{%
\def\@parse@version#1{\@parse@version@0#1}
\def\@parse@version@#1/#2/#3#4#5\@nil{%
\@parse@version@dash#1-#2-#3#4\@nil}
\def\@parse@version@dash#1-#2-#3#4#5\@nil{%
  \if\relax#2\relax\else#1\fi#2#3#4 }
}{}
\makeatother
\documentclass[aps,prd,twocolumn,titlepage,secnumarabic,nofootinbib,eqsecnum]{revtex4-2} %reprint,
\usepackage{graphicx}% Include figure files
\usepackage{dcolumn}% Align table columns on decimal point
\usepackage{bm}% bold math
\usepackage{amssymb,amsmath,amsfonts,mathrsfs}
\usepackage{graphicx}
\usepackage[dvipsnames]{xcolor}
\usepackage{verbatim}
\usepackage{epsfig}
\usepackage{subfigure}
\usepackage{color}
\usepackage{multirow}
\usepackage{comment}
\usepackage{datetime}
\usepackage{slashed}
\usepackage{ulem}
\usepackage{framed}
\usepackage{longtable}
\usepackage[mathscr]{euscript}
\usepackage{cancel}
\usepackage{tensor}
\usepackage{hyperref}
 \usepackage{mathtools}
 \usepackage[multiple]{footmisc}
 \usepackage{pgfplots}
\usepackage{natbib,bibentry}
\usepackage[applemac]{inputenc}
\def\journal@prd{prd}%

\newcommand{\nn}{\nonumber}

\def\be{\begin{equation}}
\def\ee{\end{equation}}
\def\bea{\begin{eqnarray}}
\def\eea{\end{eqnarray}}

\begin{document}

\title{Turning black-holes and D-branes \\
inside out their photon-spheres}
\author{Massimo Bianchi,}
\author{Giorgio Di Russo}

\affiliation{Dipartimento di Fisica,  Universit\`a di Roma ``Tor Vergata"  \& Sezione INFN Roma2, Via della ricerca scientifica 1, 00133, Roma, Italy}
\begin{abstract}
Very much as extremal Reissner-Nordst\"om BHs, D3-branes and their intersecting bound-states in lower dimensions enjoy a peculiar symmetry under conformal inversions that exchange the horizon with infinity and keep the photon-sphere fixed. We explore the implications of this symmetry for the dynamics of massless and massive BPS particles. In particular we find a remarkable identity between the scattering angle of a probe impinging from infinity and the in-spiralling angle of a probe with the very same energy and angular momentum falling into the horizon from inside the photon-sphere. We argue for the identity of the radial actions and Shapiro time-delays for the two processes, when some cutoff regulator is adopted. We spell out the detailed conditions for the inversion symmetry to hold in the case of large BPS BHs in four dimensions. We address conformal inversions for other BHs and Dp-branes with photon-spheres in various dimensions. We briefly discuss the fate of the symmetry at the quantum level as well as for non spherically symmetric BHs and branes and sketch potential implications for the holographic correspondence.
\end{abstract}

%
%\emailAdd{massimo.bianchi@roma2.infn.it}
%\emailAdd{alfredo.grillo@roma2.infn.it}
%
\maketitle
\tableofcontents

%%%%%%%%%%%%%%%%%%%%%%%%%%%%%%%%%%%%%%%%%%%%%%%%%%%%%%%%%%%%%%%%%%%%%%%%%%%%%%%%%%%%%%%%%%%%%%%%%
%%%%%%%%%%%%%%%%%%%%%%%%%%%%%
\section{Introduction}
%%%%%%%%%%%%%%%%%%%%%%%%%%%%%

Direct detection of gravitational waves (GWs) emitted by binary mergers of black holes (BHs) and neutron stars (NSs) has triggered increasing attention onto the detailed features of the signals that can unveil the inner characteristics of very compact gravitating objects \cite{Cardoso:2017cqb, Barack:2018yly, Cardoso:2019rvt, Barausse:2020rsu}. 

In particular, astonishing progress has been achieved in the determination of the corrections to the GW signal emitted in the ``in-spiral'' phase, relying on the connection with scattering amplitudes \cite{Bjerrum-Bohr:2018xdl, KoemansCollado:2018hss, Kalin:2019rwq, Kalin:2019inp}, as well as in the ``ring-down'' phase, relying on the peculiar connection between quasi-normal modes (QNMs) and quantum Seiberg-Witten curves for N=2 supersymmetric Yang-Mills theories \cite{Aminov:2020yma, Bianchi:2021xpr, Bonelli:2021uvf, Bianchi:2021mft}. 

Other important features, such as the presence of echoes \cite{Cardoso:2017cqb, Ikeda:2021uvc} for exotic compact objects, such as horizonless micro-states in the fuzz-ball proposal \cite{Lunin:2001jy, Mathur:2009hf, Bena:2016ypk, Bianchi:2016bgx, Bianchi:2017bxl, Bianchi:2017sds}, 
and memory effects \cite{Strominger:2014pwa, Pasterski:2015tva, Strominger:2017zoo, Addazi:2020obs, Aldi:2020qfu, Aldi:2021zhh}, based on soft theorems  \cite{He:2014laa, Cachazo:2014fwa, Bianchi:2015yta, Bianchi:2015lnw, Godazgar:2019dkh} have been investigated and precise characteristics have been identified that should allow to distinguish BHs from stringy fuzz-balls \cite{Guo:2017jmi, Bena:2020see, Bianchi:2020bxa, Bena:2020uup, Bianchi:2020miz, Mayerson:2020tpn, Bah:2021jno}. 

A common and crucial feature of BHs, D-branes and other compact gravitating objects is the `photon-sphere' or light-ring (light-halo for rotating objects) formed by the unstable bound orbits of massless particles such as photons `surfing' the wall separating the asymptotically flat region from the horizon (or the inner region for fuzz balls) \cite{Chandrasekhar:1985kt, Bianchi:2018kzy, Bianchi:2020des, Bianchi:2020yzr, Bacchini:2021fig}. 

Aim of the present investigation is to point out a remarkable symmetry between these two regions for special classes of BHs and D-branes. Indeed we will prove that
\begin{equation}
\Delta\phi_{fall}(J,E) =  \Delta\phi_{scatt}(J,E)
\end{equation}
for impact parameters $b=J/E$ above the critical value $b_c$. While $\Delta\phi_{scatt}(J,E)$ denotes the deflection angle of a massless probe with `energy' $E$ and orbital angular momentum $J$ {\bf outside} the photon-sphere, $\Delta\phi_{fall}(J,E)$ denotes the angle described by a massless object with the very same `energy' $E$ and angular momentum $J$ falling into the horizon from {\bf inside} the photon-sphere. 

Our `inversion' formula bears some resemblance with the equally remarkable ``boundary to bound" (B2B) correspondence in binary processes between periastron advance $\Delta\phi(J,E)$ and the scattering angle $\chi(J,E)$  \cite{Kalin:2019rwq, Kalin:2019inp}
\begin{equation}
\Delta\phi(J,E) = \chi(J,E) + \chi(-J,E) \qquad E<0
\end{equation}
While the latter requires analytic continuation in the binding energy $E$ and angular momentum $J$ but seems valid to all orders in the post-Minkowskian (PM) expansion for (non-rotating) compact objects in the `conservative' sector  \cite{Kalin:2019rwq, Kalin:2019inp}, the former requires no analytic continuation but seems valid only for massless or massive BPS probes in a restricted number of cases, including extremal Reissner-Nordstr\"om (RN) BHs, D3-branes, 2-charge small BHs in $d=5$ and 4-charge `large' BHs in $d=4$. 

The geometric peculiarity of these special cases is not only extremality but also their symmetry under conformal inversions of the Couch-Torrence (CT) kind \cite{CouchTorr}, that have been recently revived and generalised \cite{Cvetic:2020kwf, Cvetic:2021lss,Cvetic:1995uj} also in connection with Freudenthal duality \cite{Borsten:2018djw, Borsten:2019xas} and BH (in)stability \cite{Aretakis:2011ha, Aretakis:2011hc, Aretakis:2012ei, Godazgar:2017igz,Cvetic:2020axz}.
 
The plan of the paper is as follows. 
After briefly reviewing the concepts of critical impact parameters and photon-spheres for null geodesics in section \ref{CritNullPhotSph}, we show that the inversion formula holds true for extremal charged RN BHs in $d=4$ in Section \ref{InvSymmExtrRN}. 

In Section \ref{D3brane} we demonstrate the validity of the inversion formula for D3-branes thanks to both a homological argument based on contour deformation and a geometric argument based on a generalized Couch-Torrence inversion that keeps the photon-sphere fixed. We then pass to analyse intersecting D3-D3' `small' BHs in Section \ref{D3D3'case} and for 4-charge `large' BPS BHs in Section \ref{4QD3BH}. While the former follow very easily from the previous case any numbers of intersecting D3-D3', in the latter case we find restrictions on the four charges for the inversion symmetry to hold. 

We consider massive BPS probes in Section \ref{KKprobes} and find similar subtle restrictions on the charges and masses for 4-charge `large' BPS BHs but no such issues for 2-charge systems.  

In Section \ref{3QBHDpbrane} we find that extremal BPS 3-charge `large' non-rotating BHs in $d=5$ do not seem to enjoy the inversion formula. We also find similar difficulties for other BHs and Dp-branes with photon-spheres in various dimensions with flat or AdS asymptotics. We briefly address the issue for rotating black-holes, that will be discussed more thoroughly in \cite{MBGdRTurnHalo}. 

In Section \ref{ImplicAdSCFT} we discuss the fate of the inversion formula at the quantum level, possibly including the regularisation of the `radial' action by introducing a cutoff or putting the system in AdS. We also sketch potential implications for the AdS/CFT.

Section \ref{SummConcl} contains a summary of our results, our conclusions and directions for further investigation in the future.

 \section{Critical null geodesics and photon-spheres}
\label{CritNullPhotSph}

Scattering of (massless neutral) probes impinging from (asymptotically flat) infinity off compact rotationally-invariant (non-spinning) gravitating objects \footnote{The discussion can be easily generalised to (A)dS asymptotics with very little effort, as we will see, and with some effort to rotating objects.}\footnote{In rotating case the critical radius varies in an interval $r_H=r_{min}<r_c<r_{max}$ \cite{Chandrasekhar:1985kt, Bianchi:2018kzy, Bianchi:2020des, Bianchi:2020yzr}.}, such as BHs or D-branes, typically exposes three different regimes, depending on the value of the impact parameter $b=J/E$
\cite{Chandrasekhar:1985kt, Bianchi:2017sds}
\begin{itemize}
\item  $b>b_c$: above a critical value, the probe scatters off with a deflection angle $\Delta\phi \sim G_N M/b^{d{-}3}$ in $d$ dimensions (with $G_N$ the Newton constant and $M$ the mass) ;
\item  $b=b_c$: at the critical value, the probe is (asymptotically) trapped in an unstable `circular' orbit $r=r_c \sim b_c \sim  (G_N M)^{1/{d{-}3}}$; 
\item $b<b_c$: below the critical value, the probe falls into the horizon.
\end{itemize}

\begin{center}
\begin{figure}[!ht]
\includegraphics[scale=0.5]{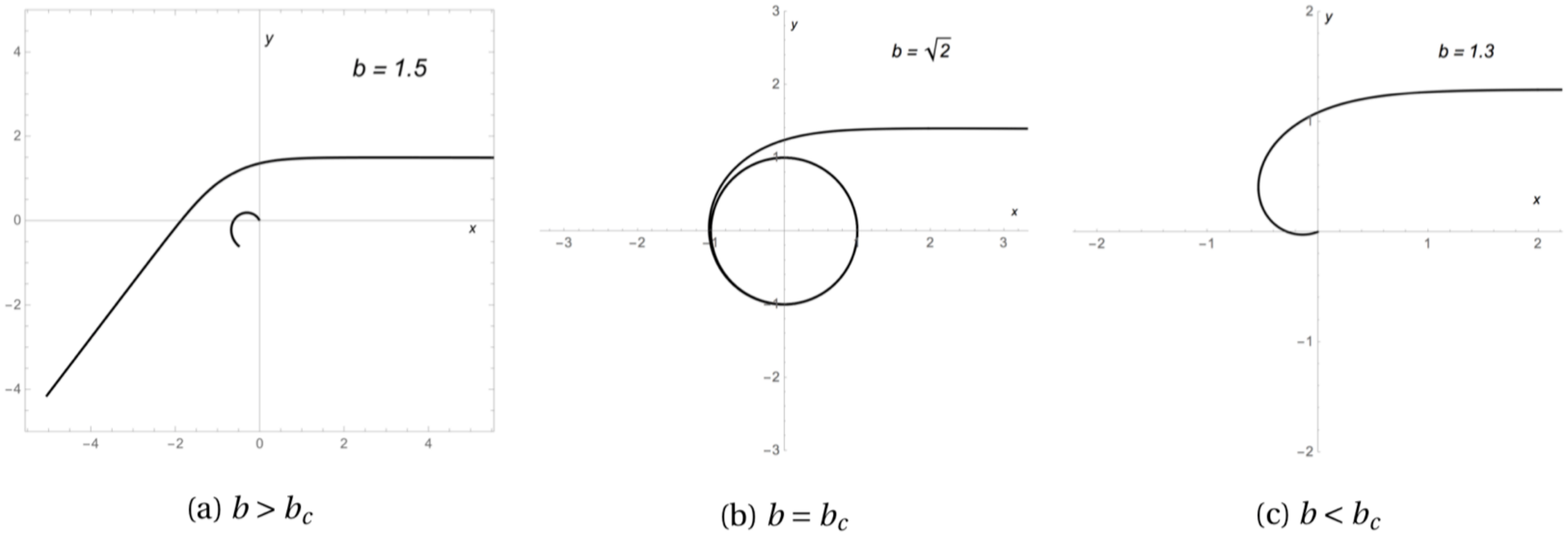}  
\caption{Geodesics in D3 brane geometry $(L=1)$ for b$>b_c$, $b=b_c$ and $b<b_c$.}\label{fig1}
\end{figure}
\end{center}

In figure \ref{fig1} we plot the geodesics for D3 branes since these will be studied in details. The critical impact parameter $b_c$ and the critical radius $r_c$ are determined by the conditions 
\begin{equation}
V_{eff}(r_c) = E^2 \quad \text{and} \quad  V'_{eff}(r_c) = 0
\end{equation}
where $V_{eff}$ is the `effective' potential.

For a spherically symmetric compact object, described by an asymptotically flat (or AdS) metric of the form
\begin{equation}
\label{ds2inr}
ds^2 = - f(r) dt^2 + {dr^2\over f(r)} + r^2 ds^2_{S^{D{-}2}}
\end{equation}
the null geodesic equation $ds^2=0$ can be separated and put in Hamiltonian form 
\begin{equation}
\label{Hinr}
{\cal H} = 0 = - {E^2\over f(r)}  + f(r) P_r^2 + {J^2\over r^2} 
\end{equation}
where
\begin{equation}
\label{conjmomenta}
E= -P_t =  f(r) \dot{t}\quad , \quad P_r =  {\dot{r}\over f(r)} \quad , \quad J= P_\phi =  r^2\dot{\phi}
\end{equation}
Thanks to spherical symmetry, only the total orbital angular momentum $J$ and its conjugate angular variable, denoted by $\phi$, play a role.
\onecolumngrid
\twocolumngrid

Extracting $P_r$ and computing the `radial' action 
\begin{align}
\label{radaction}
S_r(J,E;r_i,r_f) =& \int_{r_i}^{r_f} P_r(J,E) dr =\\\nn =&\pm\int_{r_i}^{r_f}\sqrt{E^2{-}f(r){J^2\over r^2}}{dr\over f(r)} 
\end{align}
in terms of the conserved energy $E$ and angular momentum $J$, as well as of the initial $r_i$ and final $r_f$ radii,  one can easily compute the deflection angle\footnote{We are assuming that $r_i, r_f$ are either independent of $J$ or are turning points of the radial motion so that $P_r(r_{i/f}) = 0$.}
\begin{align}
\label{Deltaphi}
\Delta\phi =& {\partial S_r(J,E;r_i,r_f)\over \partial J} = - \int_{r_i}^{r_f} {J dr \over r^2 f(r) P_r(J,E)} =\\\nn =&\int_{r_i}^{r_f} {J dr \over E \sqrt{r^4 - b^2 r^2 f(r)} }
\end{align}
and Shapiro time delay\footnote{With $r_i, r_f$ either independent of $E$ or turning points of the radial motion so that $P_r(r_{i/f}) = 0$.} 
\begin{equation}
\label{Deltat}
\Delta{t} = - {\partial S_r(J,E;r_1,r_2)\over \partial E} = - \int_{r_1}^{r_2} {E dr \over f(r)^2 P_r(J,E)} 
\end{equation}

For Dp-branes, their bound states and other extremal/BPS objects it turns out to be convenient to use a different radial variable such that $u=0$ at the horizon $r=r_H$. The metric in isotropic form reads 
\begin{equation}
\label{ds2inu}
ds^2 = - {dt^2 - d\mathbf{x}^2 \over h(u)} + h(u)\left [du^2 + u^2 ds^2_{S^{8{-}p}}\right]
\end{equation}
Setting to zero the $p$ `longitudinal' momenta $P_x=0$, as well as the  Kaluza-Klein (KK) ones, if present, one has 
 \begin{equation}
 \label{Hinu}
{\cal H} = 0 = - h(u) {E^2}  +  {1\over h(u)} \left[P_u^2 + {J^2\over u^2}\right] 
\end{equation}
so much so that the `radial' action becomes
\begin{equation}
\label{uradaction}
S_u(J,E;u_1,u_2) 
%= \int_{u_1}^{u_2} P_u(J,E) du
 = \pm \int_{u_1}^{u_2} \sqrt{ h(u)^2 E^2 -{J^2\over u^2}} du
\end{equation}
and the deflection angle and time delay turn out to be given by
\begin{equation}
\label{Deltaphitinu}
\Delta\phi = {-}\int_{u_1}^{u_2} {J du \over u^2 P_u(J,E)} \quad , \quad 
\Delta{t} = {-} \int_{u_1}^{u_2} {E h(u)^2 du \over P_u(J,E)} 
\end{equation}

%%%%%%%%%%%%%%%%%%%%%%%%%%%%%
\section{Inversion symmetry for extremal RN BHs}
\label{InvSymmExtrRN}
%%%%%%%%%%%%%%%%%%%%%%%%%%%%%%

In Einstein-Maxwell theories, the most general spherically symmetric BH in $d=4$ is the (non-extremal) RN BHs. Setting $G_N=1$, the metric is given by (\ref{ds2inr}) with
\begin{equation}
\label{fRNnonextr}
f(r)= 1- {2M\over r} + {Q^2\over r^2}
\end{equation}
For zero charge $Q=0$ one gets a Schwarzschild BH. For $|Q|=M$ one gets an extremal RN BH. The two horizons are located at 
\begin{equation}
\label{horizonsRNnonextr}
r_H^{\pm} = M \pm \sqrt{M^2-Q^2}
\end{equation}

The effective potential is given by
\begin{equation}
\label{VeffRNnonextr}
V_{eff}(r)=E^2-\dot{r}^2= {J^2\over r^2}\left[1- {2M\over r} + {Q^2\over r^2}\right]
\end{equation}
Imposing $V_{eff}=E^2$ and $V_{eff}'=0$ one finds the critical radius $r_c$ and the critical impact parameter $b_c$ \cite{Bianchi:2020des, Bianchi:2020yzr}:
\begin{equation}
\label{rbcritRNnonextr}
r_c= {1\over 2} \left(3M+\sqrt{9M^2-8Q^2}\right) 
\end{equation}
\begin{equation}
b_c = \sqrt{2 r_c^3 \over r_c{-}M} =\sqrt{\frac{27M^4{-}36M^2Q^2{+}8Q^4{+}M\left(9M^2{-}8Q^2\right)^{\frac{3}{2}}}{2\left(M^2-Q^2\right)}}
\end{equation}

\begin{figure}
%\centering
\includegraphics[width=1\linewidth]{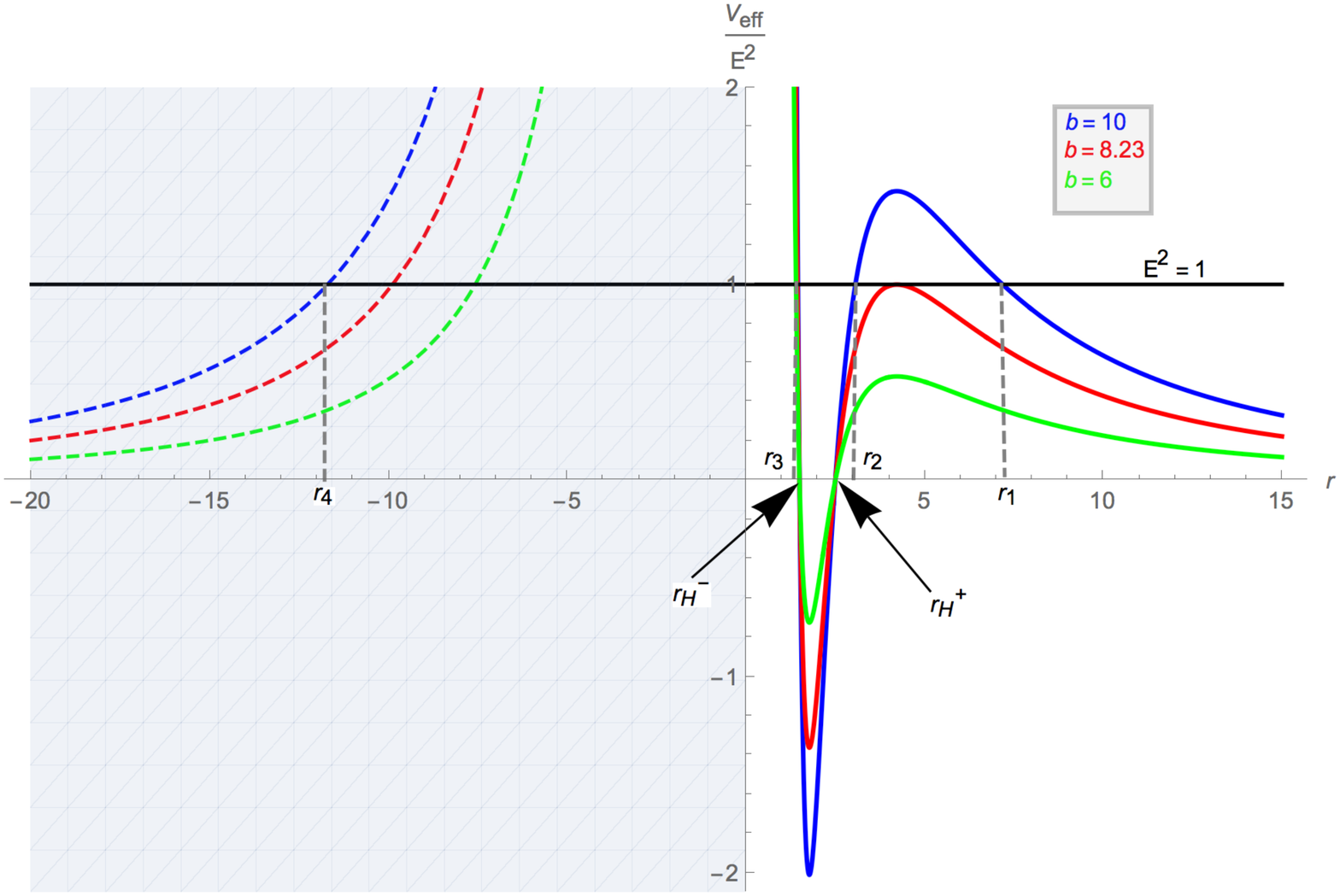}  
\caption{Non-extremal RN effective potential for $M=2$, $Q^2=3.75$.}
\end{figure}

For $b>b_c$ there are four solutions (`turning points') of $V_{eff}=E^2$ with 
$r_1>r_2>r^+_H>r^-_H>r_3>0> r_4$.

The scattering angle\footnote{Actually the `standard' scattering angle is $\Delta\theta = \pi- 2\Delta\phi_{scatt}$.} in given by (\ref{Deltaphi}) integrated from $\infty$ to the turning point $r_1$ (the largest root):
\begin{align}
\label{DeltaphiscattRNnonext}
\Delta\phi_{scatt} (E,J)=& -b\int_{\infty}^{r_1} \frac{dr}{\sqrt{r^4-b^2r^2f(r)}}  =\\\nn= \frac{2b}{\sqrt{r_{13}r_{24}}}&\mathcal{K}\left[\arcsin\sqrt{\frac{r_{24}}{r_{14}}};\sqrt{\frac{r_{14}r_{23}}{r_{13}r_{24}}}\right].
\end{align}
where $r_{ij}=r_i-r_j$ and $r_{i+}=r_i-r_H^+$ with $i,j=1,2,3,4$. If $b=b_c$ it's easy to see that $r_1=r_2$ and (\ref{DeltaphiscattRNnonext}) diverges, so the massless probe gets asymptotically trapped in a circular unstable orbit. The union of such orbits for photons (or other neutral massless probes) impinging from different directions generates the `photon-sphere'. 

In principle a massless probe with the very same energy $E$ and angular momentum $J$, such that $b>b_c$, can be `emitted' from inside the photon-sphere. Due to the strong gravitational attraction it cannot escape to infinity but rather falls into the horizon describing a `spiral'. We can compute the in-spiralling angle which is given by integrating (\ref{Deltaphi}) from $r_2$ to the horizon $r_H^+$: 
\begin{align}
\label{DeltaphifallRNnonext}
\Delta\phi_{fall} (E,J) =& -b\int_{r_2}^{r_H^+} \frac{dr}{\sqrt{r^4-b^2r^2f(r)}}   =\\\nn=\frac{2b}{\sqrt{r_{13}r_{24}}}&\mathcal{K}\left[\arcsin\sqrt{\frac{r_{13}r_{2+}}{r_{23}r_{1+}}};\sqrt{\frac{r_{23}r_{14}}{r_{13}r_{24}}}\right] .
\end{align}
In general (\ref{DeltaphiscattRNnonext}) and (\ref{DeltaphifallRNnonext}) are different for $Q\neq M$. 
In the extremal case ($Q=M$) they coincide. In terms of the coordinate $u=r-Q$, the metric reads:
\begin{align}
ds^2&=-\frac{dt^2}{h(u)}+h(u)\left(du^2+u^2d\theta^2+u^2\sin^2\theta d\phi^2\right)\\\nn h(u)&=\left(1+\frac{Q}{u}\right)^2.
\end{align}
The zeros of $P_u$ are encoded in the algebraic equation:
\begin{equation}
u^2+M(2\mp\beta)u+M^2=0
\end{equation}
where we introduced the (adimensional) impact parameter $\beta = {b\over M}=\frac{J}{M E}$. The four solutions are:
\begin{equation}\label{g}
u_{\pm}^{[\pm]}=M\left[\pm\frac{\beta}{2}-1[\pm]\sqrt{\frac{\beta^2}{4}\mp\beta}\right]
\end{equation}
where $[\pm]$ indicates the uncorrelated signs. All the solutions are real for $\beta\geq 4=\beta_c$ that corresponds to the critical impact parameter. For $\beta>4$ the ordering of the roots is the following:
\begin{equation}
u_1=u_+^+>u_2=u_+^->0>u_3=u_-^+>u_4=u_-^-
\end{equation}

Since $u_1$ is the largest root of $P_u$, the scattering angle is given by:
\begin{align}\label{10}
\Delta\phi_{scatt}(E,J)=&-b\int_{\infty}^{u_1} \frac{d u}{\sqrt{h(u)^2u^4-b^2u^2}}=\\\nn=\frac{2\beta}{\sqrt{v_{13}v_{24}}}&\mathcal{K}\left[\arcsin\sqrt{\frac{v_{24}}{v_{14}}};\sqrt{\frac{v_{23}v_{14}}{v_{13}v_{24}}}\right].
\end{align}
where we set $u_i=Mv_i$ for $i=1,2,3,4$.

Following the same procedure, we can derive the in-spiralling angle:
\begin{align}\label{12}
\Delta\phi_{fall}(E,J)=&-b\int_{u_2}^0\frac{d u}{\sqrt{h(u)^2u^4-b^2u^2}}=\\\nn=\frac{2\beta}{\sqrt{v_{13}v_{24}}}&\mathcal{K}\left[\arcsin\sqrt{\frac{v_{13}v_2}{v_{23}v_1}};\sqrt{\frac{v_{23}v_{14}}{v_{13}v_{24}}}\right].
\end{align}
and observe that the expressions (\ref{10}) and (\ref{12}) are equal if and only if
\begin{equation}\label{14}
\frac{v_{24}}{v_{14}}=\frac{v_{13}v_2}{v_{23}v_1}.
\end{equation}
that indeed holds true for $Q=M$, so much so that the angle described by the massless probe from infinity to the turning point and the in-spiralling angle are equal for extremal RN BHs:

\begin{equation}
\label{YesInversRNext}
\Delta\phi^{extrRN}_{fall} (E,J) =  \Delta\phi^{extrRN}_{scatt} (E,J).
\end{equation}

\begin{figure}
%\centering
\includegraphics[width=1\linewidth]{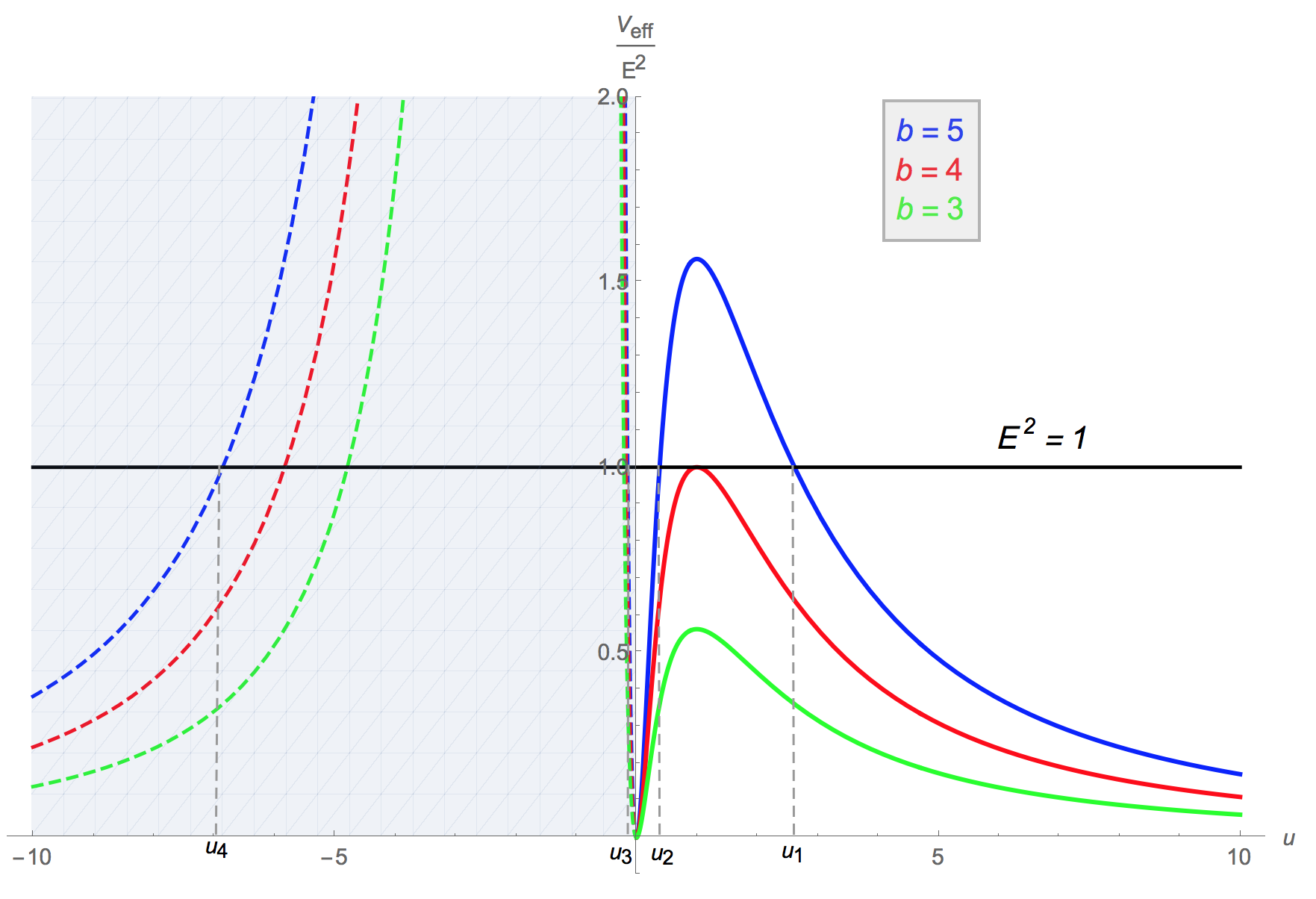}
\caption{Extremal RN effective potential $Q=M=1$. The horizon is a double zero of the metric.}
\end{figure}

We will have more to say about this remarkable relation after discussing D3-branes and their bound states. Suffice it to say here that the extremal RN metric admits a conformal inversion $u\rightarrow Q^2/u$ that preserves the light-cone ($ds^2=0 \rightarrow ds^2=0$) up an over all Weyl factor $W=u^2/Q^2$ and keeps the photon-sphere $u_c=Q$ ($r_c=2Q>r_H=Q$) fixed. This kind of transformations were introduced by Couch and Torrence \cite{CouchTorr} and recently revived by \cite{Cvetic:2020kwf, Cvetic:2021lss, Borsten:2018djw, Borsten:2019xas}.  

Before concluding this section, let us briefly consider the fate of the inversion symmetry for spherically symmetric charged BHs in AdS$_4$\footnote{In principle dS corresponds to $\ell^2\rightarrow -\ell^2$}, whereby
\begin{equation}
\label{fRNnonextr}
f(r)= 1- {2M\over r} + {Q^2\over r^{2}} + {r^2\over \ell^2} 
\end{equation}
Computing the radial momentum $P_r$ one finds 
\begin{align}
r^4 f^2&(r) P^2_r E^{-2} =r^4 - b^2 r^2 f(r) =\\\nn&=  r^4 \left(1 - {b^2 \over \ell^2}\right) - b^2 r^2 + 2Mb^2 r - Q^2 b^2 
\end{align}
Positivity of the leading term at infinity requires  $|b|<\ell$.
In the asymptotically AdS case, for $M=Q$ one finds
\begin{align}
\Delta\phi^{AdS}(b&={J/E})  = \int_{r_i}^{r_f} {b dr \over \sqrt{r^4 - b^2 r^2 f(r)} } =\\\nn&=  \Delta\phi^{flat}\left(\tilde{b} = { b/\sqrt{1 - {b^2 \over \ell^2}}}\right)  
\end{align}
As a result $\Delta\phi^{AdS}_{scatt}(J,E)=  \Delta\phi^{AdS}_{fall}(J,E)$ for charged BHs in AdS with $M=Q$ as in flat space-time\footnote{Strictly speaking the solution with $M=Q$ is a naked singularity without a proper horizon \cite{Duff:1999gh}. Regular charged BHs in AdS have non-zero angular momentum \cite{Carter:1968ks}. Yet a photon-sphere is present at $r_c=2r_0=2M=2|Q|$, independent of $\ell$, in the non-rotating case.}. Even though the radial action is not invariant under $r-r_0\rightarrow (r_c-r_0)^2/(r-r_0)$, with $r_c=2r_0=2M=2|Q|$, one expects Couch-Torrence conformal inversions \cite{CouchTorr} to admit a generalisation in gauged (super)gravity \cite{Cvetic:2020kwf, Cvetic:2021lss, Borsten:2018djw, Borsten:2019xas}.

Let us now pass to consider Dp-branes with metric (\ref{ds2inu}) where:
\begin{equation}
h(u)=H(u)^{\frac{1}{2}},\quad\quad H(u)=1+\frac{L^{7-p}}{u^{7-p}}.
\end{equation}
In particular we will mostly focus on D3-branes and their bound states.

%%%%%%%%%%%%%%%%%%%%%%%%%%%%%%%%%%
\section{D3-branes}
\label{D3brane}
%%%%%%%%%%%%%%%%%%%%%%%%%%%%%%%%%%

In the case of D3-branes, the turning points, i.e. the zeros of $h(u)$ satisfy:
\begin{equation}\label{3}
{u^4}-b^2 u^2+{L}^4=0
\end{equation}
and are thus given by
\begin{equation}\label{sold3}
u_{\pm}^2=\frac{b^2}{2}\left(1\pm\sqrt{1-\frac{4{L}^4}{b^4}}\right)
\end{equation}
so that the critical `radius' and impact parameter are $u_c=L$ and $b_c=\sqrt{2}L$ \cite{Bianchi:2017sds, Bianchi:2018kzy, Bianchi:2020des, Bianchi:2020yzr}.
%\begin{figure}
%\centering
%\includegraphics[scale=0.4]{D3_pot}
%\caption{D3-brane effective potential for $L=1$.}
%\end{figure} 
Setting
\begin{equation}
{\gamma}=\frac{\sqrt{2}{L}}{b},
\end{equation}
we can see from (\ref{sold3}) that for $0<{\gamma}<1$ we have two real positive solutions such that $u_+>u_-$ and two real negative solutions $-u_+<-u_-$. The scattering angle obtains from (\ref{Deltaphitinu}) integrating from $u=\infty$ up to $u_+$:
\begin{align}\label{5}
\Delta{\phi}_{scatt} &=-b\int_\infty^{u_+}\frac{du}{\sqrt{u^4-b^2 u^2+{L}^4}}=\\\nn&=-b\int_\infty^{u_+}\frac{du}{\sqrt{\left(u^2-u_+^2\right)\left(u^2-u_-^2\right)}}.
\end{align}
Setting $v=\frac{u_*}{u}$ one can express the integral in terms of a complete elliptic function of the first kind:
\begin{equation}\label{6}
\Delta{\phi}_{scatt}=\frac{b}{u_+}\int_0^1\frac{dv}{\sqrt{\left(1-v^2\right)\left(1-\frac{u_-^2}{u_+^2}v^2\right)}}=\frac{b}{u_+}\mathcal{K}\left[\frac{u_-^2}{u_+^2}\right].
\end{equation}
that admits a representation in terms of a gaussian Hypergeometric function
\begin{align}\label{7}
&\Delta{\phi}_{scatt} =\sqrt{\frac{2}{1+\sqrt{1-{\gamma}^4}}}\mathcal{K}\left[\frac{{\gamma}^4}{\left(1+\sqrt{1-{\gamma}^4}\right)^2}\right]=\\\nn
&=\frac{\pi}{2}\sqrt{\frac{2}{1+\sqrt{1-{\gamma}^4}}}{}_2F_1\left(\frac{1}{2},\frac{1}{2};1\Biggr\rvert\frac{{\gamma}^4}{\left(1+\sqrt{1-{\gamma}^4}\right)^2}\right) .
\end{align}
For  $0<{\gamma}<1$ the argument of the hypergeometric function is smaller than 1 and this ensures convergence of the series. For ${\gamma}=1$ ($b=b_c$) the series (\ref{7}) diverges and the massless probe gets  trapped in the photon-sphere. For ${\gamma}>1$ ($b<b_c$) it falls into the horizon hovering the photon-sphere.

The in-spiralling angle is (\ref{Deltaphitinu}) integrated from the internal turning point $u_2$, that exists for ${\gamma}<1$, to the horizon $u_H=0$:
\begin{align}\label{a}
\Delta{\phi}_{fall}&=-b\int_{u_-}^0\frac{dr}{\sqrt{u^4-b^2 u^2+{L}^4}}=\\\nn&=b\int_0^{u_-}\frac{du}{\sqrt{\left(u^2-u_+^2\right)\left(u^2-u_-^2\right)}}.
\end{align}
If we choose the new variable $u=\frac{v}{u_-}$, we can express (\ref{a}) in terms of a complete elliptic integral of the firsts kind obtaining exactly the same expression as in (\ref{6}). In other words in a D3-brane background, for impact parameters larger the critical one ($b>\sqrt{2}L$), the angle described by the massless particle coming from radial infinity and approaching the turning point is exactly equal to the in-spiralling angle.

We would like to provide two arguments to explain this peculiar and far-reaching property. The first one is a homological argument based on contour deformation for elliptic integrals. The second one is a geometric argument based on the hitherto un-noticed symmetry of the D3-brane metric under conformal inversions, generalising the ones in \cite{CouchTorr}.

\begin{figure}
%\centering
\includegraphics[width=1\linewidth]{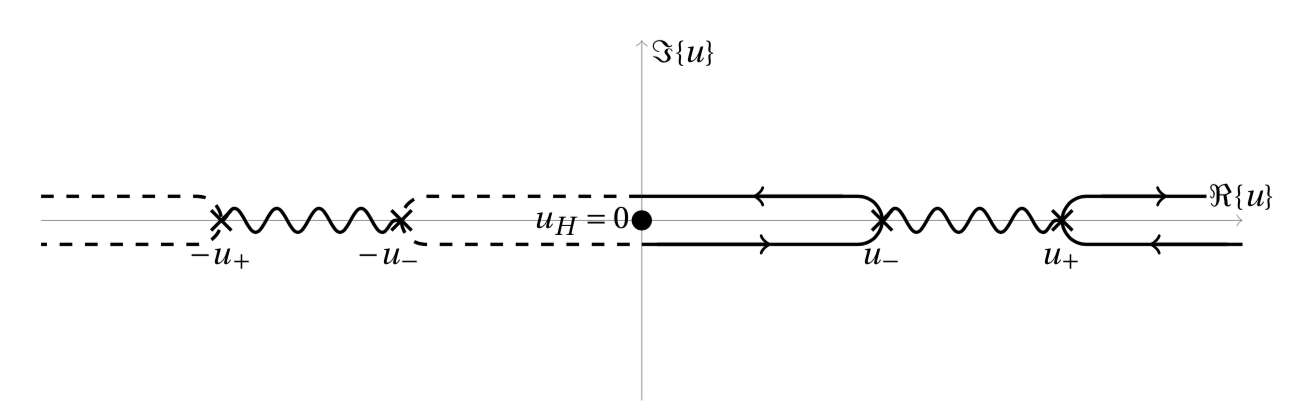}\caption{The position of the branch cut in the complex u plane.}\label{Fig1}
\end{figure}

The homological argument runs as follows. The integral $\Delta{\phi}_{fall} = {\cal I}(0,u_-)$ can be extended to negative values of $u$ so that $2\Delta{\phi}_{fall} = {\cal I}(-u_-,u_-)$, closing the contour in the ${Im}(u)<0$ half-plane one has $4\Delta{\phi}_{fall}= \oint_a \omega$  i.e. it can be viewed as the $a$-period of the elliptic curve (torus). Similarly extending beyond infinity $2\Delta{\phi}_{scatt} = {\cal I}(+u_+,-u_+)$, closing the contour in the ${Im}(u)<0$ half-plane one has $4\Delta{\phi}_{scatt}= \oint_{a'} \omega$  i.e. it can be viewed as the $a'$-period of the elliptic curve (torus). But the cycles $a$ and $a'$ are homologous so that $\Delta{\phi}_{fall}= \Delta{\phi}_{scatt}$. See figure \ref{Fig1}.

The geometric arguments runs as follows. Performing the transformations $u\rightarrow L^2/v$ the (relevant part of the) D3-brane metric suffers a Weyl rescaling 
$$
ds^2={-}\left(1 {+} {L^4\over u^4}\right)^{-{1\over 2}} dt^2+\left(1 {+} {L^4\over u^4}\right)^{-{1\over2}}\left(du^2{+}u^2d{\phi}^2\right)  \rightarrow 
$$
\begin{equation}
\quad \rightarrow {L^2\over v^2} \left[ {-} \left(1 {+} {L^4\over v^4}\right)^{-{1\over2}} dt^2+ \left(1 {+} {L^4\over v^4}\right)^{-{1\over2}}\left(dv^2{+}v^2d{\phi}^2\right) \right]
\end{equation}
Since massless geodesics are Weyl invariant and $u_-= L^2/u_+$ the integrals corresponding to the two angles get exchanged under conformal inversion $u\rightarrow L^2/u$, whose fixed locus is the photon-sphere $u_c=L$. This is a remarkable symmetry of D3-branes that -- if not spoiled by quantum corrections -- can reveal new insights into the holographic AdS/CFT correspondence. 
We will see that the same property is enjoyed by intersecting D3-branes systems with two and four charges. In the latter case this is related to generalised Freudenthal duality \cite{Cvetic:2020kwf, Cvetic:2021lss, Borsten:2018djw, Borsten:2019xas}. After imposing suitable restrictions on the charges, it amounts to a generalised Couch-Torrence inversion. 

\section{D3-D3' `small' BHs}
\label{D3D3'case}

Let us consider a massless probe in an intersecting D3-D3' background\footnote{This system is T-dual to the D1-D5 system. We work in the D3-D3' U-duality frame to keep a uniform notation.}  compactified on $T^4\times S^1$. We denote by $y$ the coordinate  compactified on a (large) circle $S^1$, by $\vec{x}$ the coordinates along the four (1-4) non compact spatial (DD) directions and the by $\vec{z}$ the coordinates along the four (ND) directions (6-9) compactified on a (small) $T^4$. After smearing each D3's along the transverse $T^2$ directions, the metric is given by:
\begin{align}\label{gdr1}
ds^2=&- \left(H_3H_{3'}\right)^{-{1\over2}}(dt^2-dy^2)+\left(H_3H_{3'}\right)^{{1\over2}}dx^2+\\\nn&+\left(\frac{H_3}{H_{3'}}\right)^{{1\over2}}ds_{T^2_{67}}^2 + \left(\frac{H_{3'}}{H_3}\right)^{1/2}ds_{T^2_{89}}^2
\end{align}

with

\begin{equation}
H_3(u)=1+\frac{L_3^2}{u^2}\quad,\quad H_{3'}(u)=1+\frac{L_{3'}^2}{u^2}
\end{equation}
Setting
\begin{equation}
x_1+ix_2=u\cos\theta e^{i\psi}\quad , \quad  x_3+ix_4=u\sin\theta e^{i\phi}.
\end{equation}
the 4-dimensional metric in the DD directions reads
\begin{equation}
dx^2 = du^2 + u^2 [ d\theta^2 + \sin^2\theta d\phi^2 + \cos^2\theta d\psi^2 ]
\end{equation}
The zero mass shell condition Hamiltonian formalism reads:
\begin{align}
\mathcal{H}=&\sqrt{H}(P_y^2{-}P_t^2)+\frac{1}{\sqrt{H}} \left[{P_u^2}{+}\frac{P_\theta^2}{u^2}{+}\frac{P_\phi^2}{u^2\sin^2\theta}{+}\frac{P_\psi^2}{u^2\cos^2\theta}\right]{+}\\\nn+&|\vec{P}_z|^2\sqrt{H_3\over H_{3'}} + |\vec{P}'_z|^2\sqrt{H_{3'}\over H_{3}}=0
\end{align}
where $H=H_3H_{3'}$. 

Thanks to spherical symmetry, without loss of generality, one can consider motion in the equatorial plane ($\theta=\pi/2$). As a result $J_\psi =0$ and $J_\phi = J$ is the total angular momentum. For simplicity, for the time being, we also take vanishing KK momenta $\vec{P}_z=\vec{P}_z'=0$ along the compact directions. We will  consider massive BPS probes later on.  

The only difference between D3 and intersecting D3-D3' lies in the definition of the (harmonic) function $H(u)$. 
Indeed we expect D3-D3' systems to enjoy the same property that characterizes D3-branes.
In the present case the turning points satisfy
\begin{equation}\label{c}
\left(1+\frac{L_{3}^2}{u^2}\right)\left(1+\frac{L_{3'}^2}{u^2}\right)\frac{u^4}{b^2}-u^2=0
\end{equation}
whose solutions are:
\begin{equation}
u_{\pm}^2=\frac{1}{2} \left\{b^2-L_{3}^2-L_{3'}^2\pm
\sqrt{(b^2-L_{3}^2-L_{3'}^2)^2-4L_{3}^2L_{3'}^2} \right\}
\end{equation}

Analogously to D3 case we can compute the scattering angle, described by the probe from infinity to the `external' turning point $u=u_+$, as well as the in-spiralling angle, from the `internal' turning point $u=u_-$ to the horizon $u_H=0$. Taking into account the difference between roots (\ref{3}) and (\ref{c}) the integrations can be performed in an identical way and the formulae obtained for the scattering and the spiralling angle are the same:

\begin{equation}\label{d}
\Delta\phi_{scatt}(\infty,u_+)=\Delta\phi_{fall}(u_-,0)=\frac{b}{u_+}\mathcal{K}\left[\frac{u_-^2}{u_+^2}\right]
\end{equation}
where $u_\pm$ in (\ref{d}) are given by (\ref{c}).

%\begin{figure}
%\centering
%\includegraphics[scale=0.4]{D3D3_pot}
%\caption{D3D3'-brane system effective potential  for $L_{3'}=1$ and $L_{3'}=2$. }
%\end{figure} 

Once again the conformal inversion symmetry
\begin{equation}
u\rightarrow {L_{3}L_{3'}\over u} 
\end{equation}
under which $u_+ = L_{3}L_{3'}/u_-$ and the photon-sphere $u_{c} =\sqrt{L_{3} L_{3'}}$ is fixed, is 
crucial to explain the geometric origin of the result. 

We do not repeat here the homological argument based on contour deformation, because it runs exactly the same way as for D3-branes. 

%%%%%%%%%%%%%%%%%%%%%%%%%%%%%
\section{Intersecting D3-branes as large BPS BHs}
\label{4QD3BH}
%%%%%%%%%%%%%%%%%%%%%%%%%%%%%

Intersecting four stacks of D3-branes, such that any pair has 4 common N-D (internal) directions, one gets a `large' BPS BH solution in STU supergravity with four charges $Q_i$. Neglecting the internal $T^6$, to which we will turn our attention later on, the (4-d) metric is given by
\begin{align}\label{e}
ds^2=&-\prod_{i=1}^4\left(1+\frac{Q_i}{u}\right)^{-{1\over2}}dt^2+\\\nn&+\prod_{i=1}^4\left(1+\frac{Q_i}{u}\right)^{{1\over2}}\left[du^2{+}u^2\left(d\theta^2+\sin^2\theta d\phi^2\right)\right]
\end{align}
Our aim is to prove that the property that the scattering angle and the in-spiral angle holds true for all backgrounds of the form (\ref{e}). The case in which all the charges are equal coincides with an extremal RN background that we already dealt with in Section (\ref{InvSymmExtrRN}). We first identify the most general set of charges that lead to solutions admitting a conformal inversion symmetry \`a la Couch-Torrence.
Then  we will analyze the case $Q_1=Q_2>Q_3=Q_4$ that still admits an analytical treatment and briefly discuss the unequal charge case at the end. 

%%%%%%%%%%%%%%%%%%%%%%%%%%%%%
\subsection{Couch-Torrence conformal inversion}
%%%%%%%%%%%%%%%%%%%%%%%%%%%%%

In the case of 4-charge BHs obtained from intersecting D3-branes the relevant inversion is
\begin{equation}
u \rightarrow {\sqrt{Q_1Q_2Q_3Q_4}\over u}
\end{equation}
This is a conformal isometry of the metric if and only if\footnote{Otherwise one has to perform a (symplectic) transformation on the charges as well \cite{Cvetic:2020kwf, Cvetic:2021lss, Borsten:2018djw, Borsten:2019xas}.}
\begin{equation}
\sum_i Q_i = \sqrt{Q_1Q_2Q_3Q_4} \sum_i {1\over Q_i}
\end{equation}
This happens to be the case when all $Q$'s are equal or when they are equal in pairs. Setting 
\begin{equation}
Q_1 = x Q_4 \quad , \quad Q_2 = y Q_4 \quad , \quad Q_3= z Q_4
\end{equation}
the condition boils down to
\begin{equation}
1+x+y+z = \sqrt{xyz} \left( 1+ {1\over x} +{1\over y} +{1\over z}\right)
\end{equation}

In general setting $z=\lambda^2 xy$ one finds that the only three solutions are 
\begin{equation}
\lambda^2 = 1  \quad , \quad  \lambda^2 = {x^2 \over y^2} \quad , \quad  \lambda^2 = {y^2 \over x^2}
\end{equation}
that in turn mean  (assuming all charges to be positive)
\begin{equation}
z = xy  \quad , \quad  x=yz \quad , \quad  y=zx
\end{equation}
or even more simply 
\begin{equation}
Q_1Q_2 = Q_3Q_4 
\end{equation}
or permutations thereof! It is amusing to see that the simplest non trivial integer solution is (a permutation of) $Q_1=1$, $Q_2=2$, $Q_3=3$, $Q_4=6$. In general we have a three-parameter family of solutions admitting conformal inversion as a symmetry. In all these cases the photon-sphere, located at
\begin{equation}
u_c = \sqrt[4]{Q_1Q_2 Q_3Q_4}
\end{equation}
is fixed under inversion and $u_1u_2 = u_c^2= \sqrt{Q_1Q_2 Q_3Q_4}$. As a consequence the identity $\Delta\phi_{scatt} = \Delta\phi_{fall}$  has a deep geometric origin that allow to turn these 4-charge BHs inside out their photon-spheres. 

\subsection{Pair-wise equal charges}

For simplicity we set $Q_1=Q_2=Q$ and $Q_3=Q_4=\tilde{Q}$ and we choose $Q>\tilde{Q}$. The turning points satisfy:

\begin{equation}\label{f}
{\left(1+\frac{Q}{u}\right)^2\left(1+\frac{\tilde{Q}}{u}\right)^2}-{b^2}u^2=0.
\end{equation}

Setting:

\begin{equation}
z=\frac{u}{Q}\hspace{1cm}\beta=\frac{b}{Q}\hspace{1cm}q=\frac{\tilde{Q}}{Q}<1,
\end{equation}

the square root of (\ref{f}) reads:

\begin{equation}
(1+z)(q+z)=\pm \beta z
\end{equation}

which admits the following solutions:

\begin{equation}\label{h}
z_\pm^{[ \pm ]}=\frac{-(q+1\mp\beta)[\pm]\sqrt{(q-1)^2+\beta^2\mp2\beta(q+1)}}{2}
\end{equation}

where the square parenthesis means the uncorrelated signs. Let's notice that in the limit $q\rightarrow 1$ we recover the roots (\ref{g}) that allows us to order the roots in (\ref{h}) as follows:

\begin{equation}\label{20}
z_1=z_+^+>z_2=z_+^->0>z_3=z_-^+>z_-^-=z_4.
\end{equation}

The structure of the scattering and in-spiralling angle is the same as in (\ref{10}) and (\ref{12}) and, as for extremal RN, the roots (\ref{h}) satisfy a relation analogous to (\ref{14}).

\begin{figure}
%\centering
\includegraphics[width=1\linewidth]{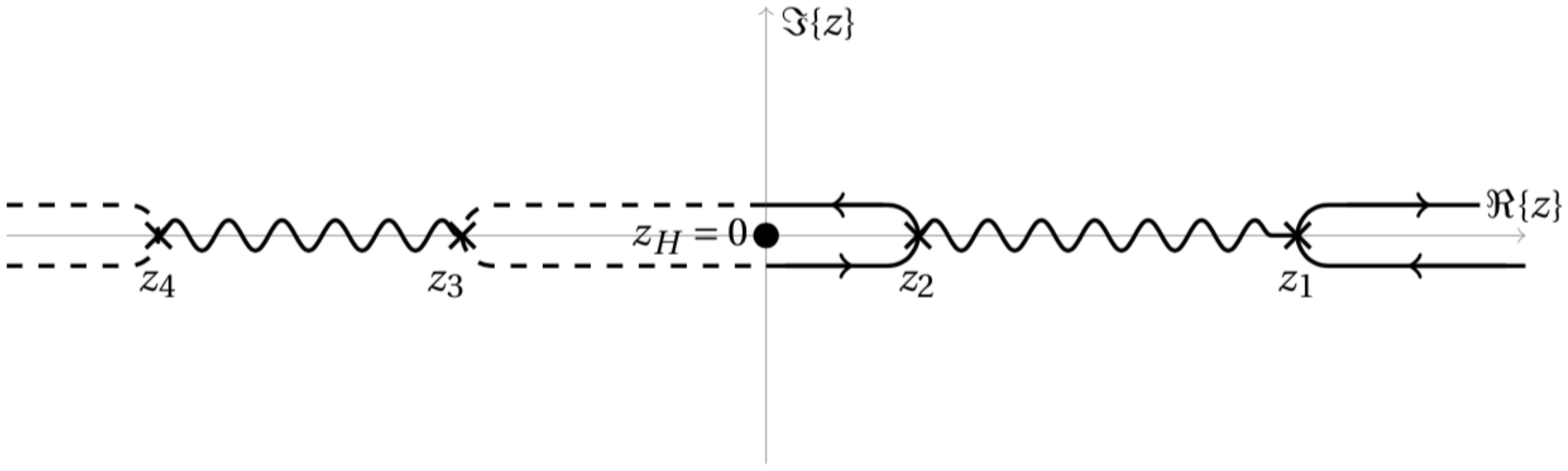}
\caption{The position of the branch cuts in the complex $z$ plane}
\end{figure} 

\subsection{All unequal charges}

In general, for different charges the effective potential for massless particles is given by:

\begin{equation}
V_{eff}(u)= \frac{J^2}{u^2\prod_{i=1}^4\left(1+\frac{Q_i}{u}\right)}
\end{equation}

Setting 
\begin{equation}
z=\frac{u}{Q_4}\hspace{1cm}\beta=\frac{b}{Q_4}\hspace{1cm}q_i=\frac{Q_i}{Q_4},\quad i=1,2,3
\end{equation}
without loss of generality one can take $0<q_i<q_j<1$ with $i<j=1,2,3$. The effective potential has a minimum in $z=0$ and a maximum in $z=z_c>0$. The critical impact parameter is identified by the following relations:
\begin{align}
&V_{eff}(z=z_c)=E^2 \quad,\quad V_{eff}'(z=z_c)=0\\\nn&\beta_c=\frac{\sqrt{(1+z_c)(q_2+z_c)(q_3+z_c)(q_4+z_c)}}{z_c}
\end{align}

\begin{figure}
%\centering
\includegraphics[width=1\linewidth]{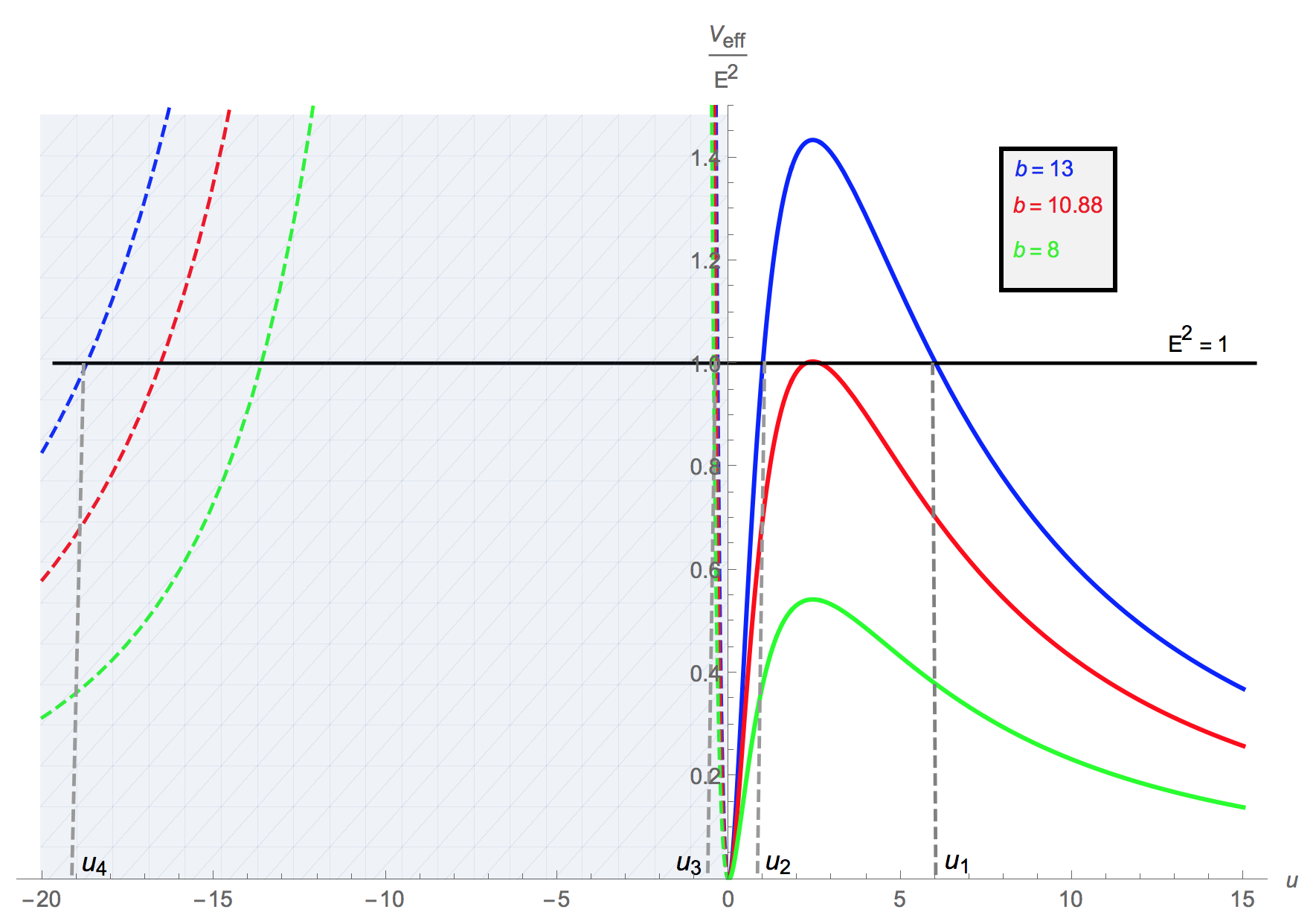}
\caption{D3D3D3D3-brane effective potential  for $Q_1=1$, $Q_2=2$, $Q_3=3$ and $Q_4=4$.}
\end{figure} 

We are interested in the regime in which $\beta>\beta_c$. The zeros of $P_u$ are encoded in the following algebraic equation:
\begin{equation}\label{22}
(1+z)(q_1+z)(q_2+z)(q_3+z)=\beta^2z^2
\end{equation}
The expressions for the scattering and in-spiralling angles are the same as in (\ref{10}) and (\ref{12}), where now the $z_i$ are the solutions of (\ref{22}).  

Although (\ref{22}) can be solved by quadrature, the expressions for the $z_i$ are quite cumbersome to manipulate. In order to check that even in this case
\begin{equation}
\Delta\phi_{scatt}^{4Q} = \Delta\phi_{fall}^{4Q}
\end{equation}
holds true it turns out convenient to use numerical methods that confirm indeed the validity of our inversion formula, provided $Q_1Q_2=Q_3Q_4$ or permutations thereof. 

One may try and extend the analysis to 4-charge STU BHs in AdS with metric \cite{Duff:1999gh}  
\begin{align}\label{AdSBHmetric}
ds^2=&{-} f(u) \prod_{i=1}^4H_i^{-{1\over2}}dt^2+\\\nn &\prod_{i=1}^4H_i^{{1\over2}}\left[{du^2\over f(u)}{+}u^2\left(d\theta^2{+}\sin^2\theta d\phi^2\right)\right]
\end{align}
with $H_i = 1+ {Q_i\over u}$ and $f = 1 + {r^2\over \ell^2} \prod_{i=1}^4H_i$. It is easy to check that 
\begin{equation}
\Delta\phi_{AdS}^{4Q} (b=J/E) = \Delta\phi_{fall}^{4Q}\left(
\tilde{b}={b\over \sqrt{1-{b^2\over \ell^2}}}
\right)
\end{equation} 
for generic choice of the charges $Q_i$ so much so that 
\begin{equation}
\Delta\phi_{AdS}^{4Q, scatt} (J,E) = \Delta\phi_{AdS}^{4Q, fall} (J,E)
\end{equation}
if $Q_1Q_2=Q_3Q_4$ or permutations thereof. Yet, as already mentioned in the case of singly charged BHs with $M=Q$ in AdS, strictly speaking \eqref{AdSBHmetric} has a naked singularity and no proper horizon \cite{Duff:1999gh}, even though a photon-sphere at the same  $u=u_c$ as in flat space-time is present that is fixed under CT transformations $u\rightarrow u_c^2/u$, exchanging infinity and the `putative' horizon at $u=0$. Regular BHs with arbitrary charge in AdS require non-zero angular momentum \cite{Carter:1968ks}.

\section{Massive BPS probes}
\label{KKprobes}

So far we have only considered massless probes. Let us try and consider some massive probe. For simplicity we will focus on massive BPS particles that owe their mass to their (generalised) KK momentum along internal directions. Since in the case of D3-branes there are none, we will consider intersecting D3-branes.
For non BPS particles with arbitrary masses and couplings to the geometry we do not expect the identity to hold.  

\subsection{D3-D3 with massive BPS probes}

In the intersecting D3 brane configuration (\ref{gdr1}) with $L_3=L_{3'}=L$, one can consider the hyper-plane $\theta = \pi/2$ without loss of generality. The relevant part of the metric reads:

\begin{equation}\label{gdr2}
ds^2=H^{-\frac{1}{2}}\left(-dt^2+dy^2\right)+H^{1/2}(du^2+u^2d\theta^2)+d\mathbf{z}^2.
\end{equation}
where $H=H_3 H_{3'}=h^2$ with $h=1 +{L^2\over u^2}$.

The mass shell condition $\mathcal{H}=g^{MN}p_Mp_N=0$ holds exactly in 10 dimensions, but we allow non-zero momenta associated to the internal $\mathbf{z}$ coordinates. In the Hamiltonian formalism, imposing the mass-shell condition yields
\begin{equation}\label{gdr4}
P_u^2=h^2(u)\mathcal{E}^2-m^2h(u)-\frac{J^2}{u^2},\quad\mathcal{E}^2=E^2-p_y^2,\quad m^2= |\mathbf{p_z}|^2 \quad .
\end{equation}
The effective potential, the massive particle is subject to, is given by
\begin{equation}\label{gdr3}
\frac{V_{eff}(u)}{\mathcal{E}^2}=\frac{u^2}{u^2+L^2}\left(\mu^2+\frac{b^2}{u^2+L^2}\right)\quad,\quad\mu=\frac{m}{\mathcal{E}}.
\end{equation}
The radial turning points are:
\begin{equation}\label{gdr5}
u^2_\pm=\frac{\lambda^2}{\nu}(1\pm\sqrt{1-\nu^2}),
\end{equation}
with
\be
\nu=\frac{2L^2\sqrt{1-\mu^2}}{b^2+\mu^2L^2-2L^2},\hspace{1cm}\lambda=\frac{L}{\sqrt[4]{1-\mu^2}}
\ee
Obviously $\mu^2<1$ and we are interested in the regime in which $\nu<1$. The effective potential (\ref{gdr3}) is symmetric under exchange $u\rightarrow -u$ and this fact is reflected in the position of the roots on real axis:
\begin{equation}
u_+>u_->0>-u_->-u_+ .
\end{equation}
The expressions for scattering and in-spiralling angles are respectively:
\begin{equation}
\Delta\phi_{scatt}=-\beta\int_{\infty}^{u_+}\frac{du}{\sqrt{(u^2-u_+^2)(u^2-u_-^2)}}
\end{equation}
\begin{equation}
\Delta\phi_{fall}=-\beta\int_{u_-}^{0}\frac{du}{\sqrt{(u^2-u_+^2)(u^2-u_-^2)}}
\end{equation}
where $\beta=b/\sqrt{1-\mu^2}$. It is crucial to note that the roots of (\ref{gdr5}) obey $u_1u_2=\lambda^2$, so starting from the scattering angle and performing the coordinate transformation $v=\frac{u_1u_2}{u}=\frac{\lambda^2}{u}$, that leaves the photon-sphere at $u_c (L, \mu) = \lambda$ fixed, it is very easy to demonstrate that $\Delta\phi(\infty,u_+)=\Delta\phi(u_-,0)$.

The generalisation to $L_{3}\neq L_{3'}$ is straightforward. We only write down the expression for the 'radial' momentum 
\begin{align}\label{boh1}
P_u^2=H_3H_{3'}\mathcal{E}^2-&m^2H_3 - m'^2H_{3'}-\frac{J^2}{u^2} = \\\nn
= \mathcal{E}^2 \left(1{-}\mu^2{-}\mu'^2\right)& \left[  1 {+} {L^2\left(1{-}\mu^2\right) {+} L'^2 \left(1{-}\mu'^2\right) {-} b^2 \over \left(1{-}\mu^2{-}\mu'^2\right)u^2} \right.\\\nn
\left.+ {L^2L'^2\over \left(1-\mu^2-\mu'^2\right) u^4} \right]
\end{align}
with $m^2 = |\mathbf{p}|^2 = \mu^2\mathcal{E}^2$ and $m'^2 = |\mathbf{p}'|^2 = \mu'^2\mathcal{E}^2$. The angular deflection is given by
\begin{equation}\label{boh2}
\Delta\phi = \tilde{b} \int_{u_i}^{u_f} {du \over \sqrt{u^4 + \left(\tilde\alpha^2 - \tilde{b}^2\right)u^2 + \lambda^4}}
\end{equation}
with 
\begin{align}
&\tilde{b} {=} b /\sqrt{1{-}\mu^2{-}\mu'^2}\quad , \quad \tilde\alpha^2 = {L^2\left(1{-}\mu^2\right) {+} L'^2 \left(1{-}\mu'^2\right) \over 1{-}\mu^2{-}\mu'^2}\quad ,\\\nn &\lambda^2 = LL' /\sqrt{1{-}\mu^2{-}\mu'^2}
\end{align}
The turning points are 
\begin{equation}\label{boh3}
u^2_\pm = {1\over 2} \left[ \tilde{b}^2 - \tilde\alpha^2 \pm \sqrt{ \left(\tilde{b}^2 - \tilde\alpha^2\right)^2 - 4 \lambda^4}\right]
\end{equation}
so that the critical impact parameter is given by
\begin{equation}\label{boh4}
\tilde{b} = \sqrt{\tilde\alpha^2 + 2 \lambda^2}
\end{equation}
while the photon-sphere is located at 
\begin{equation}\label{boh5}
\tilde{u}_c= \lambda = {\sqrt{LL'} \over \sqrt[4]{1-\mu^2-\mu'^2}}
\end{equation}
Using the homological argument or the conformal inversion $u\rightarrow \lambda^2 / u$ one easily proves the identity
\begin{equation}\label{boh6}
\Delta\phi^{mKK}_{scatt} = \Delta\phi^{mKK}_{fall}
\end{equation}
for generic KK masses and D3 and D3' charges.

\subsection{D3-D3-D3-D3 with massive BPS probes}

The generalisation to 4-d BHs with 4 charges associated to 4 stacks of intersecting D3-branes is subtler. For six generic KK momenta $\mathbf{p}_{ij}=\mathbf{p}_{ji}$ with $i\neq j$, $i,j=1,... 4$, such that $m_{ij}^2 = |\mathbf{p}_{ij}|^2 = \mu_{ij}^2E^2$, satisfy the 10-d mass-shell condition, the radial momentum is given by
\begin{align}\label{boh7}
P_u^2=&\mathcal{E}^2\prod_{i=1}^4 H_i(u)  - \sum_{i<j}^6 m_{ij}^2 H_iH_j - \frac{J^2}{u^2} =\\\nn &=\mathcal{E}^2 \left(1-\mu^2\right) \left[ 1 + {\sigma_1 \over u} + {\sigma_2 - \tilde{b}^2 \over u^2} + {\sigma_3 \over u^3} +  {\sigma_4 \over u^4} \right]
\end{align}
where $\mu^2 = \sum_{i<j} \mu_{ij}^2$, $\tilde{b} = b/\sqrt{1-\mu^2}$ 
and
\begin{equation}
\sigma_1 = {\sum_i Q_i \left(1{-}\sum_{j\neq i} \mu_{ij}^2\right)  \over 1-\mu^2}
\quad , \quad 
\sigma_2 = {\sum_{i<j} Q_i Q_j (1{-}\mu_{ij}^2)  \over 1-\mu^2} 
\end{equation}
\begin{equation}
\sigma_3 = {\sum_{i<j<k} Q_i Q_j Q_k \over 1-\mu^2} 
\quad , \quad
\sigma_4 = {Q_1Q_2Q_3Q_4 \over 1-\mu^2}
\end{equation} 

Conformal inversions of the Couch-Torrence kind correspond to
\begin{equation}
u\rightarrow {\sqrt{\sigma_4} \over u}
\end{equation}  
this is a symmetry of the metric if and only if
\begin{equation}\label{90}
\sigma_1 \sqrt{\sigma_4} = \sigma_3
\end{equation} 
which is a non-trivial constraint on $Q_i$ and $m_{ij}$, whose solution, up to permutations, is 
\begin{equation}\label{XX}
Q_1= x Q_4 \quad Q_2= y Q_4 \quad Q_3= \lambda^2 xy Q_4 
\end{equation}
with
\begin{equation}\label{XX}
\mu_{12}^2 = \mu^2  \quad \lambda = \sqrt{1-\mu^2} \quad \mu^2_{ij} =0 \quad \text{for} \quad {(i,j)}\neq (1,2)
\end{equation}
or 
\begin{equation}\label{XX}
\mu_{34}^2 = \mu^2  \quad \lambda = {1\over \sqrt{1-\mu^2}}\quad \mu^2_{ij} =0 \quad \text{for} \quad {(i,j)}\neq (3,4)
\end{equation}

In these cases (constrained charges and KK momentum) the photon-sphere is located at $u_c= \sqrt[4]{\sigma_4}$ and it is easy to check that  
\begin{equation}
\Delta\phi^{KK}_{scatt} = \Delta\phi^{KK}_{fall}
\end{equation}
either by algebraic or numerical means. 

However, for non BPS particles with arbitrary masses and couplings to the geometry we do not expect the identity to hold.

%%%%%%%%%%%%%%%%%%%%%%%%%%%%%
\section{ Higher-dimensional BHs and branes}
\label{3QBHDpbrane}

%%%%%%%%%%%%%%%%%%%%%%%%%%%%%

After the success obtained for D3-branes and intersecting D3-brane systems, it seems quite natural to inquire whether BHs and branes in higher dimensions that expose a photon-sphere admit a similar inversion symmetry. We anticipate that the answer is negative. The basic reason is the very different behaviour of the geometry at infinity from the geometry at the horizon. 

Nevertheless we briefly analyse the 5-dimensional non-rotating case and even more briefly sketch the generalisation to higher dimensions and AdS asymptotics.    

\subsection{Non-rotating (BPS) BHs in 5-dimensions}

In type IIB compactifications on $T^5$, 5-dimensional non-rotating BHs with nonzero horizon area can be constructed by superposing $Q_5$ D5-branes, $Q_1$ D1-branes and Kaluza-Klein momentum $Q_p$ . The $Q_5$ D5-branes are wrapped on $T^5$. The $Q_1$ D-strings are wrapped along one of the directions of the torus and the KK momentum $P = N/R$ along the string. The solution is given in terms of three harmonic functions $H_1$, $H_5$ and $H_P$.

\begin{equation}
H_1=1+\frac{Q_1}{u^2},\hspace{1cm}H_5=1+\frac{Q_5}{u^2},\hspace{1cm}H_p=1+\frac{Q_p}{u^2}
\end{equation}
with $u^2=x_1^2+...+x_4^2$. The metric reduced to 5 dimensions in spherical coordinates is:
\begin{align}
ds_5^2&=-\frac{dt^2}{\left(H_1H_5H_p\right)^{\frac{2}{3}}}+\left(H_1H_5H_p\right)^{\frac{1}{3}}\left[du^2+\right.\\\nn&\left.+u^2\left(d\theta^2+\sin^2\theta d\phi^2+\cos^2\theta d\psi^2\right)\right].
\end{align}
In the hyper-plane $\theta=\pi/2$, 
%\begin{equation}
%\mathcal{L}=\frac{1}{2}\left[-\frac{\dot{t}^2}{H^{2/3}}+H^{1/3}\left(\dot{u}^2+u^2\dot{\phi}^2\right)\right]
%\end{equation}
setting $H=H_1H_5H_p$, the Hamiltonian for massless probes can be written as
%
%The canonical conjugate momenta are:
%\begin{align}
%&p_t=\frac{\partial \mathcal{L}}{\partial \dot{t}}=-\frac{\dot{t}}{H^{2/3}}=-E\\
%&p_r=\frac{\partial \mathcal{L}}{\partial \dot{r}}=H^{1/3}\dot{r}\label{a1}\\
%&p_\phi=H^{1/3}r^2\dot{\phi}=J\label{a2}
%\end{align}
\begin{equation}\label{a3}
0=\mathcal{H}=\frac{1}{2}\left[-H^{2/3}E^2+H^{-1/3}\left(P_u^2+\frac{J^2}{u^2}\right)\right].
\end{equation}
If, for simplicity, we consider $Q_1=Q_5=Q_p=Q$, the zeros of $P_u$ satisfy
\begin{equation}\label{a5}
u^6+(3Q-b^2)u^4+3Q^2u^2+Q^3=0.
\end{equation}
Setting $\zeta=u^2/Q, \beta^2=b^2/Q$, (\ref{a5}) becomes:
\begin{equation}\label{a6}
\zeta^3+(3-\beta^2)\zeta^2+3\zeta+1=0
\end{equation}
This third degree equation has three real roots $\zeta_i, i=1,2,3$ only for $\beta>3\sqrt{3}/2$, which are such that:
\begin{equation}
\zeta_1>\zeta_2>0>\zeta_3.
\end{equation}

The angle described by the particle coming from infinity and reaching the turning point $u_1=\sqrt{Q\zeta_1}$ is given by
\begin{equation}
\label{a8}
\phi(\infty,r_1){=}
{-}b\int_\infty^{u_1}\frac{rdr}{\sqrt{(u^2{-}Q\zeta_1)(u^2{-}Q\zeta_2)(u^2{-}Q\zeta_3)}}\stackrel{u^2=x}{=}
\end{equation}
$$
\quad  -\frac{b}{2}\int_\infty^{u_1^2=Q\zeta_1}\frac{dx}{\sqrt{(x{-}Q\zeta_1)(x{-}Q\zeta_2)(x{-}Q\zeta_3)}}
\stackrel{Q\zeta_1/ x=\xi}{=} 
$$
$$
\quad \frac{b}{2\sqrt{Q}}\sqrt{\frac{-\zeta_1}{\zeta_2\zeta_3}}\int_0^1\frac{d\xi}{\sqrt{\xi(\xi-1)\left(\xi-\frac{\zeta_1}{\zeta_2}\right)\left(\xi-\frac{\zeta_1}{\zeta_3}\right)}}$$
Since $\zeta_1/\zeta_2>1>0>\zeta_1/\zeta_3$, the last integral in (\ref{a8}) can be written in terms of the complete elliptic integral of the first kind:
\begin{equation}\label{a10}
\phi(\infty,u_1)=\frac{b}{\sqrt{Q\zeta_{13}}}\mathcal{K}\left[\sqrt{\frac{\zeta_{23}}{\zeta_{13}}}\right].
\end{equation}
The in-spiralling angle can be computed in a similar way:
\begin{equation}\label{a9}
\phi(u_2,0)=-b\int_{u_2}^0\frac{udu}{\sqrt{(u^2{-}Q\zeta_1)(u^2{-}Q\zeta_2)(u^2{-}Q\zeta_3)}}\stackrel{u^2=x}{=}
\end{equation}
$$
=-\frac{b}{2}\int_{u^2_2}^0\frac{dx}{\sqrt{(x-Q\zeta_1)(x-Q\zeta_2)(x-Q\zeta_3)}}\stackrel{\xi=x/u_2^2}{=}
$$
$$
=\frac{b}{2u_2}\int_0^1\frac{d\xi}{\sqrt{\left(\xi-\frac{\zeta_1}{\zeta_2}\right)(\xi-1)\left(\xi-\frac{\zeta_3}{\zeta_2}\right)}}.
$$

Since $\zeta_1/\zeta_2>1>0>\zeta_3/\zeta_2$, the last integral in (\ref{a9}) can be written in terms of incomplete elliptic integral:

\begin{equation}\label{a11}
\phi(u_2,0)=\frac{b}{\sqrt{Q\zeta_{13}}}\mathcal{K}\left[\arcsin\sqrt{\frac{\zeta_2\zeta_{13}}{\zeta_1\zeta_{23}}};\sqrt{\frac{\zeta_{23}}{\zeta_{13}}}\right]
\end{equation}

Notice that (\ref{a10}) and (\ref{a11}) are different in general and cannot be rendered equal for any choice of $\beta$. The deep reason of the inequality is the lack of symmetry between horizon $u_H=0$ and infinity $u\rightarrow \infty$. While the latter is a branching point the former is a regular point. Moreover as we discuss more extensively momentarily, generalised Freudenthal duality in $D=5$ exchanges particles with strings.  

\subsection{Other p-branes}

The relevant formula (under square root in the denominator) {\it viz.}
\begin{equation}
u^4 P^2_u = F(u) = u^4H_p(u) - u^2 b^2
\end{equation}
suggests absence of a photon-sphere (critical geodesics) for $p\ge 5$ and its presence for $p\le 4$. We already discussed at length the case $p=3$. Let us consider the other cases with $p\le 4$, for which
\begin{equation}
F(u) = u^4H_p(u) - u^2 b^2 = u^4 + L^{7-p} u^{p-3} - b^2 u^2 
\end{equation}
so that
\begin{equation}
u_c = \sqrt[7{-}p]{5-p \over 2}  L_p  \quad , \quad b_c = \sqrt{7{-}p \over 5-p} u_c\end{equation}
and 
\begin{equation}
\Delta\phi = \int {b du \over \sqrt{u^4 + L^{7-p} u^{p-3} - b^2 u^2}}
\end{equation}
More explicitly one has 
\begin{equation}
{p=4} \quad : \quad  \Delta\phi = \int {b du \over \sqrt{u^4 + L^3 u - b^2 u^2}}
\end{equation}
\begin{equation}
{p=2} \quad : \quad  \Delta\phi = \int {b du \sqrt{u}\over \sqrt{u^5 + {L^5} - b^2 u^3}}
\end{equation}
\begin{align}
{p=1} \quad : \quad  \Delta\phi &= \int {b udu \over \sqrt{u^6 + L^6 - b^2 u^4}} =\\\nn&= \int {b d\xi \over 2 \sqrt{\xi^3 + L^6 - b^2 \xi^2}} 
\end{align}
\begin{equation}
{p=0} \quad : \quad  \Delta\phi = \int {b u^{3/2} du \over \sqrt{u^7 + {L^7} - b^2 u^5}}
\end{equation} 
In all of the above integrals the behaviour at the horizon $u_H=0$ is very different from the one at infinity $u\rightarrow \infty$ and no obvious inversion symmetry can be envisaged that exchange the two and keeps the photon-sphere fixed. At present we cannot exclude a symmetry under a generalised inversion such as $u\rightarrow L^{1+a} u^{-a}$ with $a$ not an integer, that would however be non-involutive.  

One argument that should help explaining this problem is the fact that Freudenthal duality would exchange point-particles with strings in $D=5$ \cite{Borsten:2018djw, Borsten:2019xas}. Generalized
Freudenthal duality should exchange $p$-branes with $D{-}4{-}p$-branes. This is why self-dual objects like particles / BHs in $D=4$, D3 in $D=10$ or strings in $D=6$ enjoy this property\footnote{The case of D2, that admit a photon-sphere as we have aleeady seen, in $D=8$ is subtler since no solution of the form $AdS_4\times S^4 \times {\cal M}$ with ${\cal M}$ some compact manifold (such as $T^2$ for strings o $T^3$ for M-theory) seems to be known. The best one can do is
$AdS_4\times CP^3$  in Type IIA or $AdS_4\times S^7/Z_k$ in M-theory. We thank M.~Trigiante for pointing this out.}.

 \section{Eikonal phase and radial action}
\label{ImplicAdSCFT}

Let us now discuss possible implications of our classical geodetic analysis for scattering amplitudes in a putative quantum theory of gravity, such as string theory. 
In the eikonal limit, valid for large impact parameters $b>b_c$, the scattering amplitude of a (massless) probe off a spherically symmetric target is given by the exponential of the eikonal phase \cite{DAppollonio:2010krb, Bianchi:2011se, DAppollonio:2015fly,  Kulaxizi:2018dxo, Kulaxizi:2019tkd,  Parnachev:2020zbr}
\begin{equation}
\widetilde{\cal S}(b,E) = 1+i\widetilde{\cal T}(b,E) = 1+ i{\widetilde{\cal A}(b,E)\over 2E}\approx e^{2i\delta_{eik}(J=bE,E)}
\end{equation}
where $\widetilde{\cal A}(b,E)$ is the scattering amplitude in impact parameter space
\begin{equation}
\widetilde{\cal A}(b,E)= \int {d^{d-2}q\over (2\pi)^{d-2}} e^{i\vec{q}{\cdot}\vec{b}} {\cal A}(\vec{q},E)
\end{equation}
with $\vec{q}$ the transferred (space-like) momentum. In turn $\delta_{eik}$ can be written in terms of the `radial' action
\begin{align}
\delta_{eik}(J,E) &\approx S_r(J,E; r_i, r_f) = \int_{r_i}^{r_f} P_r(J,E) dr =\\\nn
&=\int_{r_i}^{r_f} \sqrt{E^2 - f(r) {J^2\over r^2}} {dr \over f(r)}
\end{align}
Observables such as the deflection angle and the time delay are then given as derivatives of $\delta_{eik}$, {\it viz.}
\begin{equation}
\Delta\phi(J,E) = {\partial \delta_{eik} \over \partial J} \quad , \quad \Delta{t}(J,E)  = - {\partial \delta_{eik} \over \partial E}
\end{equation}
When $r_i$ or $r_f$ are taken to infinity or to the horizon there might be divergences that can be subtracted or regulated by introducing a boundary/wall such as in AdS.

Indeed in asymptotically flat space-times $P_r \approx \sqrt{E}$ at very large $r$ and
$S_r\approx \sqrt{E} R$ is linearly divergent with the cutoff $R$ as expected for a nearly free particle. Yet, even in this case, the deflection angle $\Delta\phi$ remains finite for $b<b_c$ and diverges only for $b=b_c$ {\it i.e.} for critical geodesics. This is not the case for  $\Delta{t}$ that obviously diverges, unless the process takes place in an asymptotically AdS (aAdS) \cite{Kulaxizi:2018dxo, Kulaxizi:2019tkd,  Parnachev:2020zbr}. The raising term $r^2/\ell^2$ in $f(r)$ regulates the integral for $J< \ell E$ {\it i.e.} for $b<\ell$ as well as it derivatives.

Whether in aAdS or not, the (regulated) radial action is the crucial ingredient in the semi-classical dynamics of massless or massive probes in presence of BHs, D-branes or other gravitating objects. Thanks to the identity $\Delta\phi_{scatt} (J,E)= \Delta\phi_{fall}(J,E)$, that extends to $S_r(J,E; r_1, \infty) = S_r(J,E; r_2, r_H)$ when both are finite or up to subtractions, one can explore near-horizon dynamics by performing experiments in the asymptotically flat region. 

Let us stress once again that the identity relies on generalised inversions \`a la Couch-Torrence \cite{CouchTorr}, that is a conformal symmetries of the metric for D3-branes, D3-D3' and for 4-d `large' BPS BHs with four charges, that satisfy the condition $Q_1Q_2=Q_3Q_4$ or permutations thereof, such that the photon-sphere is the fixed locus of conformal inversion. Moreover it extends {\it mutatis mutandis} to massive (BPS) probes whose geodetic equations are not conformal invariant. 

Another story is the fate of the inversion symmetry at the quantum level. Since it acts by conformal transformations of the metric it is likely to be anomalous. Yet, being an element of the U-duality group in special cases (when e.g. $\sum Q= \sqrt{\prod Q} \sum Q^{-1}$) it may survive in a full quantum theory of gravity as string theory. In this context, as already mentioned, it may help by-passing issues of extrapolating the results for large $b>>b_c$ to small $b<b_c$ since physics at the horizon may be captured by physics at flat infinity thanks to the remarkable property we found.

\section{Summary and conclusions}
\label{SummConcl}

Let us summarise the results of our investigation and identify some lines for future study.

We have shown that many BPS systems admitting a photon-sphere enjoy a peculiar symmetry under conformal inversions that keep the photon-sphere fixed and exchange horizon and flat infinity. Since the dynamics of 'massless' probes in backgrounds of this form is Weyl invariant, we have found that the scattering angle
$\Delta\phi_{scatt}(J,E)$ for a probe impinging from infinity and scattering off the compact gravitating center for $b>b_c$ exactly coincides with the in-spiralling angle
$\Delta\phi_{fall}(J,E)$ for a probe emitted from inside the photon-sphere and falling into the horizon with the very same energy $E$ and angular momentum $J$. Despite
similarity with the {\it B2B} formula \cite{Kalin:2019rwq, Kalin:2019inp} relating periastron advance to scattering angle, we should stress once again that the latter requires an analytic continuation to negative $E$, while in our case $E$ is positive and thus measurable.

Playing with numbers one can formally increase the mass $m$ of the probe but the validity of our analysis would be jeopardise. Yet for EMRI (extreme mass-ratio in spirals) where $m<<M$ our analysis is reliable and may shed some light on the highly non-linear merging phase that can only be tackled by numerical methods at present.

On astrophysical grounds, the obvious limitation is the BPS nature of the systems we have analysed. This reflects in their charge(s) and the absence of angular momentum.
In higher dimensions, {\it i.e.} $d\ge 5$, rotating BHs are compatible with BPS conditions and we plan to further investigate this issue despite the lack of a simple inversion symmetry already for (BPS) non-rotating BHs in $d\ge 5$. Yet the presence of a photon-sphere or rather a photon-halo (with $u_c$ varying in some interval depending on $b_c$) suggests that one should try and find a way to explore its interior (up to the horizon) by some generalised inversion that go beyond Couch-Torrence inversions, whereby $u\rightarrow u_c^2(b)/u$ depends on $E$ and $J$ along the lines of \cite{Cvetic:2020kwf}.

Even more intriguing is fate of the identity for horizonless objects or fuzz-balls such as 2-charge micro-states or JMaRT solution, including their BPS limit (GMS solution). It is tempting to conjecture that the relevant `inversion' keeping the photon-sphere (or rather photon-halo in these cases) fixed should exchange infinity with the `regular' origin or cap. Similar issues are raised by the study of QNMs \cite{Aminov:2020yma, Bianchi:2021xpr, Bonelli:2021uvf, Bianchi:2021mft} that, needless to say, crucially depend on the photon-sphere. We plan to study the inversion properties for waves, quantum particles and strings in the near future.

\section*{Acknowledgements}
We would like to thank A.~Amariti, G.~Bonelli, G.~Bossard, D.~Consoli, D.~Fioraventi, F.~Fucito, A.~Grillo, J.~F.~Morales, R.~Poghosyan, R.~Porto, G.~Pradisi, F.~Riccioni, R.~Russo, R.~Savelli, M.~Trigiante and A.~Tanzini for useful discussions and suggestions.

\bibliographystyle{ieeetr}
\bibliography{turningBH}

\end{document}